\documentclass[preprint2]{aastex}

\newcommand {\hii}{H\,{\sc ii}} 
\newcommand {\kms}{\relax \ifmmode {\,\rm km\,s}^{-1}\else \,km~s$^{-1}$\fi}
\newcommand {\ha}{H$\alpha$}

\newcommand {\oiii}{[O\,{\sc iii}]}
\newcommand {\nii}{[N\,{\sc ii}]}
\newcommand {\sii}{[S\,{\sc ii}]}

\shorttitle{Bubbles of N44}
\shortauthors{Naz\'e et al.}

\begin{document}

\title{Structure and Dynamics of Candidate O Star Bubbles in N44}

\author{Ya\"el Naz\'e\altaffilmark{1,5},
  You-Hua Chu\altaffilmark{2,}\altaffilmark{6},
  Mart\'{\i}n A. Guerrero\altaffilmark{2,6},
  M. S. Oey\altaffilmark{3}, 
  Robert A. Gruendl\altaffilmark{2},
  R. Chris Smith\altaffilmark{4}
}
\altaffiltext{1}{Institut d'Astrophysique et de G\'eophysique, All\'ee du 6 Ao\^ut 17, Bat. B5c, B 4000 Li\`ege (Sart-Tilman), Belgium; naze@astro.ulg.ac.be}
\altaffiltext{2}{Astronomy Department, University of Illinois, 1002 W. Green Street, Urbana, IL 61801, USA; chu@astro.uiuc.edu,  mar@astro.uiuc.edu, gruendl@astro.uiuc.edu}
\altaffiltext{3}{Lowell Observatory, 1400 West Mars Hill Rd., Flagstaff, AZ 86001, USA; oey@lowell.edu}
\altaffiltext{4}{Cerro Tololo Inter-American Observatory, Casilla 603, La Serena, Chile; csmith@noao.edu}
\altaffiltext{5}{Research Fellow FNRS (Belgium)}
\altaffiltext{6}{Visiting astronomer, Cerro Tololo Inter-American Observatory}

\begin{abstract} 

Dynamical studies of superbubbles and Wolf-Rayet ring nebulae 
show discrepancies from the standard, adiabatic model for
wind-blown bubbles.  We therefore study the physical properties and
kinematics of three candidate bubbles blown by single O stars,
to evaluate whether these discrepancies are also found in these
simpler objects.  Our sample candidates are N44\,F, N44\,J, and N44\,M, 
in the outskirts of the \hii\ complex N44 in the Large Magellanic
Cloud. We have obtained ground-based and {\it HST} emission-line
images and high dispersion echelle spectra for these objects.
From the \ha\ luminosities and the \oiii/\ha\ ratios of these nebulae, 
we estimate the spectral types of the ionizing stars to be O7V, O9.5V 
and O9.5V for N44\,F, N44\,J, and N44\,M, respectively.  We find that 
the observed expansion velocity of 12 $\rm km\ s^{-1}$ for N44\,F is
consistent with the stellar wind luminosity expected from the central
ionizing star, as predicted by the standard bubble model.  The observed
upper limits for the expansion velocities of N44\,J and N44\,M are also
compatible with the expected values, within the uncertainties.
We also report the discovery in N44\,F of strongly-defined dust
columns, similar to those seen in the Eagle Nebula.  The photoevaporation 
of these dense dust features may be kinematically important and may 
actually govern the evolution of the shell. The inclusion of photoevaporation 
processes may thus undermine the apparent agreement between the observed 
bubble dynamics and the simple adiabatic models.

\end{abstract}

\section{Introduction}

Mechanical feedback from massive stars is a fundamental
driver of galaxy evolution.  The supernovae (SNe) and supersonic stellar
winds from these stars generate bubbles and shells in the interstellar
medium (ISM), which may dominate the ISM structure formation in
many star-forming galaxies.  The standard, adiabatic model for these 
shells and superbubbles \citep[e.g.,][]{Wetal77} predicts the production 
of hot, $10^6 - 10^7$ K, low-density gas within these shells, which bear 
the products
of SN and massive star nucleosynthesis.  Therefore, these bubble structures
are thought to be the source of the diffuse, hot ionized component of
the ISM.  Vigorous star-forming regions may generate supergiant shells
that blow out of galaxy disks, thereby dispersing metals and mass far
from the parent star formation event.  In addition, this mechanical
energy contributes to ISM kinematics and is thought to be a primary
source of turbulence \citep[e.g.,][]{NF96,Go00}.

The effects of mechanical feedback have been clearly demonstrated, for 
example, in starburst galaxies \citep{Jetal00,Setal00} and in superbubbles 
around OB associations \citep[e.g.,][]{Setal92,Oey96}. However, several 
troubling discrepancies remain unsolved.  For example, almost all clean 
examples of bubbles blown by Wolf-Rayet stars and of superbubbles present 
shells that are too small for their observed parent stellar population 
\citep[e.g.,][]{TC82,Detal95}.  Many superbubbles also show anomalous 
kinematics and X-ray emission \cite[e.g.,][]{ro82,CM90,WH91}.
To resolve these puzzles and more clearly understand the mechanical
feedback process, it is therefore necessary to examine simpler systems.

In active star-forming 
regions, high concentrations of OB stars collectively produce 
superbubbles. But in their peripheries, where stars are loosely 
distributed,  discrete bubbles may be produced by individual stars.
\ha\ images of \hii\ regions ionized by OB associations in the 
Large Magellanic Cloud (LMC) have revealed large (50--150 pc) gas 
shells around the OB associations and numerous small ($<$ 15 pc), 
ring-like nebulae outside the large shell \citep[see, e.g.,][] {DEM}.
These small nebulae are candidate wind-blown bubbles of single O or B
stars \citep{WD98}, and offer perhaps the simplest test of the 
standard, adiabatic bubble model.

We have investigated the kinematics and physical nature of three
ring-like nebulae in N44 \citep{hen56} in the LMC.  N44 is a
bright \hii\ complex with a 44~pc$\times$67~pc superbubble at the
center and several compact \hii\ regions along
the shell rim.  In the surrounding area, small dense \hii\ regions as well 
as extended diffuse nebulae are also present at distances up to 160 pc.
Figure~\ref{n44tot} shows these features and their identifications 
given by \citet{hen56}.
Three OB associations exist in N44 \citep{LH70}: LH~47 in the central 
shell, LH~48 in the compact \hii\ region N44\,I, and LH~49 partially
embedded in the bright \hii\ region N44\,D.  The three nebulae
that we have studied are N44\,F, N44\,J, and N44\,M.  N44\,F is a
bright, circular \hii\ region at the northwest rim of the superbubble
and is adjacent to a bright filament connected to the main nebula. 
In contrast, N44\,J and N44\,M, appearing respectively on the  northern and 
eastern parts of N44, are fainter and relatively isolated.
None of these three nebulae have been studied since their original
identification by \citet{hen56}.

We have obtained ground-based CCD images and high-dispersion echelle 
spectra of N44\,F, N44\,J, and N44\,M to study the morphology and kinematics
of these nebulae.  Additionally, we have used {\it Hubble Space Telescope} 
({\it HST}) WFPC2 images of N44\,F in the \ha\ and \sii\ lines to analyze its
ionization structure.  In this paper, we first describe these observations 
in \S 2, then present the analysis and results for N44\,F in \S 3 and N44\,J 
and N44\,M in \S 4. We present our conclusions in \S 5.

\section{Observations} 
The datasets used in this study include: (1) CCD images taken with the 
Curtis Schmidt and 0.9~m telescopes at Cerro Tololo Inter-American 
Observatory (CTIO), (2) CCD images taken with the {\it HST} WFPC2, 
and (3) high-dispersion echelle spectra taken with the CTIO 4~m 
telescope.  The high-dispersion spectra are useful to diagnose 
expanding shells that are not morphologically identifiable. 

\subsection{Ground-based CCD Images}
CCD images of N44 were taken with the Curtis Schmidt telescope at 
CTIO in 1993 December. The detector was a front-illuminated Thomson 
1024$\times$1028 CCD camera with 19~$\mu$m pixels, or
1\farcs835~pixel$^{-1}$. Images were obtained with \ha\ 
($\lambda_c$=6561\AA, $\Delta\lambda$=26\AA), \oiii\ ($\lambda_c
$=5010\AA, $\Delta\lambda$=50\AA), and \sii\ ($\lambda_c$=6720\AA, 
$\Delta\lambda$=50\AA) filters for exposure times of 1800~s, 2700~s, 
and 2700~s, respectively.
The \ha\ filter includes not only the \ha\ line but also the 
neighboring \nii\ $\lambda\lambda$6548,6583 lines.
At the LMC's radial velocity, $\sim$300 \kms, the \nii\ $\lambda$6583 
line is red-shifted further away from the red edge of filter, but the 
\nii\ $\lambda$6548 line is red-shifted further into the \ha\ bandpass.
Using the \nii/\ha\ intensity ratio measured from our echelle 
spectra (see Section 2.3), we estimate that the \nii\ $\lambda$6548 line 
emission contributes 2.4\%, 4.3\% and 3.3\% of the total flux of N44\,F, 
N44\,J and N44\,M, respectively. This \nii\ contamination has been corrected
in our \ha\ flux measurements.

Higher-resolution \ha\ images were taken with the 0.9~m telescope 
at CTIO in 2000 December. The detector was a thinned, 
AR coated SITe 2048$\times$2048 CCD with 24~$\mu$m pixels, or
0\farcs4~pixel$^{-1}$. The \ha\ filter ($\lambda_c$=6559\AA, 
$\Delta\lambda$=64\AA) used transmits \ha\ and both \nii\ lines,
therefore the ``\ha'' images are really \ha+\nii\ images, and thus are
only used for morphology analysis. The angular resolution, determined 
by the FWHM of the stellar images in the field of view, was 
approximately 1\farcs3. To map the entire N44 complex, five fields
were observed with exposure times of 2$\times$600~s for each field. 

All CCD images were bias-subtracted and flat-fielded in a standard way 
using IRAF software, and multiple frames were combined to remove cosmic 
rays and improve the S/N ratio. Figure~\ref{n44tot} shows the \ha\ mosaic 
of the N44 complex, and Fig.~\ref{n44jm} 
presents close-ups of N44\,F, N44\,J and N44\,M in the \ha, \oiii, and \sii\ lines.

\subsection{ {\em HST} WFPC2 Images}
{\em HST} WFPC2 images of N44\,F were acquired via the Hubble 
Observation Problem Report \footnote{The original observations of N44 
were made with an incorrect orientation, resulting in a loss of 1/3 of 
the field of view. To compensate the loss, HOPR observations were granted. 
To facilitate the scheduling of these HOPR observations and to maximize 
the scientific yield, a different pointing center in N44 suitable for arbitrary 
orientations was chosen.} (HOPR) observations of the Cycle 6 program 6698. 
In these observations, made with the $F656N$ filter for 2$\times$500~s 
and the $F673N$ filter for 2$\times$600~s, N44\,F was included in the 
Wide-Field Camera 3 (WF3). The $F656N$ filter, centered at 6563.7 \AA\ 
with a FWHM of 21.4 \AA, includes the \ha\ line and the neighboring 
\nii\ lines. Using the \nii/\ha\ intensity ratio measured from our echelle 
spectra (see Section 2.3), we estimate that the \nii\ $\lambda$6548 line 
emission contributes 2.4\% of the flux in N44\,F.  This \nii\ 
contamination has been corrected in our \ha\ flux measurements. 
The $F673N$ filter, centered at 6732.2 \AA\ with a FWHM of 47.2 \AA\, 
includes the nebular lines of \sii\ $\lambda\lambda$ 6716,6731. 

The calibrated WFPC2 images were produced by the standard {\it HST} pipeline.
We processed them further with IRAF and STSDAS routines.  The images
taken with the same filter were combined to remove cosmic rays and to
produce a total-exposure map.  The combined \sii\ and \ha\ images were
then each corrected for the intensity- and position-dependent charge
transfer efficiency as explained by \citet{Ho95}.  Following the 
recommended procedures for narrowband WFPC2 photometry\footnote{available at
http://www.stsci.edu/instruments/wfpc2/Wfpc2\_faq/wfpc2\_nrw\_phot\_faq.html}, 
we find the rectangular widths of the $F656N$ and $F673N$ filters to be 28.3 
\AA\ and 63.3 \AA, respectively, and use these widths to
convert the flux densities into fluxes.

Figure~\ref{n44f} shows the WFPC2 images of N44\,F. To show the excitation 
variations, the \sii/\ha\ ratio map of N44\,F is also presented, in addition 
to the \ha\ and \sii\ images.

\subsection{CTIO 4~m Echelle Spectra} 

Echelle spectra were obtained with the CTIO 4 m telescope in 2000 December.  
The spectrograph was used with a 79 line mm$^{-1}$ echelle grating and 
the long-focus red camera.  We observed a single order centered on the 
\ha\ line by inserting a post-slit \ha\ interference filter and replacing
the cross-dispersing grating with a flat mirror.  A Tek 2048$\times$2048
CCD with 24 $\mu$m pixel$^{-1}$ was used to record the images. 
 This provided a spectral sampling of $\sim$0.08 \AA\ pixel$^{-1}$ and 
a spatial sampling of 0\farcs26 pixel$^{-1}$.  
The wavelength coverage, limited by the \ha\ filter and the echelle
order, was 125 \AA; the spatial coverage, limited by the optics,
was $\sim$200$''$.  The angular resolution, determined by 
the FWHM of the seeing, was approximately 1\arcsec.  A slitwidth of
1\farcs65 was used.  The resultant spectral resolution, measured from
the Th-Ar lamp lines, was about 13 \kms\ FWHM. 

Each of the three nebulae, N44\,F, N44\,J, and N44\,M, was observed with 
an east-west oriented slit. The exposure times were 600~s for N44\,F, 
but only 300~s for N44\,J and N44\,M. The spectral lines detected include 
nebular \ha\ and \nii\ lines, and telluric \ha\ and OH lines \citep{Oetal96}.
We have adopted rest wavelengths of 6562.7885 \AA, and 6548.080 \AA\ and 
6583.454 \AA\ for the \ha\ and \nii\ $\lambda\lambda$6548,6583 nebular 
lines, respectively \citep{Spyr95}. 

\section{Physical Structure of N44\,F} 
We will first describe the overall morphology of N44\,F, examine its 
excitation and ionization structure, and derive its rms electron 
density. Special emphasis is given to the prominent dust ``pillars''
where star formation might be taking place. Next, we describe the 
kinematic features detected in the echelle spectra, fit Gaussian 
components to the line profiles, and use the resultant velocities 
to identify expanding structures and to determine expansion velocities.  
Finally, we combine the morphological and kinematic information and 
present an integrated view of this \hii\ region. Throughout this paper,
we will adopt a distance of 50~kpc to the LMC \citep{fea}.

\subsection{Morphology and Density} 
N44\,F is a bright circular nebula nestled on a bright filament
at the northwest rim of the superbubble in N44. In low-resolution 
\ha\ images, N44\,F appears to have a double-ring structure with a 
very bright region in the northwest. 
The details of these features are resolved in the {\it HST} WFPC2 images. 
The outer ring, of radius $\sim$ 22\arcsec, is surrounded by dusty features.
Finger- or pillar-like dust columns, similar to those in the Eagle Nebula 
\citep[=M16;][]{hes96}, are also seen at the periphery, with two especially prominent 
ones at the west, close to the brightest region of N44\,F.  These dust columns 
are revealed in detail by the {\it HST} images, and 
will be discussed more extensively in Section 3.2.  The inner ring, of 
radius $\sim$ 6\arcsec, is resolved into diffuse emission without sharp 
features. The western part of this ring merges into the brightest 
region of N44\,F. A group of bright stars are located within the central 
ring, but they are substantially offset towards the bright emission region
at the northwest. Unfortunately, none of the stars in the vicinity of N44\,F 
have been studied photometrically or spectroscopically. 

The \sii\ images of N44\,F show a very different morphology: only the outer
ring is visible and it is mostly composed of sharp filaments. 
The pillar-like features are also present and stars at the
tips of two of them are clearly seen (see Fig.~\ref{pillars}). The \sii/\ha\ 
ratio in the central region is very low, about 0.1. The outer 
ring possesses a higher \sii/\ha\ ratio, with values
ranging from 0.2 to 0.4.  Enhanced \sii/\ha\ ratios are often associated 
with shell structures or high velocity shocks.  As we do not find the latter
 in the nebula (see Section 3.3), the somewhat elevated \sii/\ha\ ratio
is likely to result from the low ionization parameter\footnote{The ionizing 
flux derived further can be used to estimate the ionization 
parameter $U\varpropto(Q(H^0)\,n\,f^2)^{1/3}$, where $Q(H^0)$ is the ionizing 
flux in photons s$^{-1}$, $n$ the density in cm$^{-3}$, and $f$ the filling 
factor. Using $Q(H^0)$= 1.0$\times$10$^{49}$ photons~s$^{-1}$,
$n$=40~cm$^{-3}$ and $f$=$1-\left( 1-\frac{\Delta R}{R} \right)^3\sim$0.6,
we find a value of $\sim10^{-3}$ for the ionization parameter.} caused by the
shell morphology.  The values of \sii/\ha\ are in fact similar to those seen
in other shell structures \citep[e.g.,][]{Las77,Hun94}.

Using the Curtis Schmidt images, we also examine the \oiii\ morphology of 
N44\,F. Because of the poorer resolution, only the overall morphology can be
described. The \oiii\ image of N44\,F looks very similar to that in \ha, except 
for the bright \ha\ outer ring, which disappears almost completely in \oiii.
We have used the {\it HST} WFPC2 observations of N44\,C \citep{gar} to 
calibrate the observed \oiii/\ha\ ratio derived with the Curtis Schmidt 
images. The \oiii/\ha\ ratio of N44\,F is relatively high, about 0.5--1.2
inside the ring, with the highest value near the star(s).
The outer ring has the lowest \oiii/\ha\ ratio, about 0.25. The integrated 
\oiii/\ha\ ratio for the whole nebula is about 0.55. The high ratio near 
the central region indicates that the ionizing source must be an early-type 
massive star. 

The WFPC2 \ha\ image can be further analyzed to estimate the rms density 
of N44\,F. A color excess of E(B$-$V)=0.11~mag for N44 \citep{OM95} is
adopted in the extinction correction. The emission measure (EM) derived 
from the extinction-corrected surface brightness \citep{peim} ranges 
from 7$\times$10$^3$~cm$^{-6}$~pc in the faintest interior regions of 
N44\,F to 4$\times$10$^4$~cm$^{-6}$~pc at the tips of the ``pillars''.
To derive the density from the EM, we can assume two simple geometries. 
First, the density can be estimated by assuming the depth to be equal to
the diameter of the nebula: for the central region, we find a value
ranging from 25~cm$^{-3}$ to 35~cm$^{-3}$. Second, N44\,F is 
limb-brightened and we can therefore estimate the density by assuming a 
constant density spherical shell with radius $R$ and thickness $\Delta R$.
This geometry predicts a unique radial surface brightness profile for any
$\Delta R/R$ based on the varying path length through the shell material.
Comparing these simple modeled profiles to the observed \ha\ surface 
brightness profiles near the shell rim, we find a shell thickness 
$\Delta R/R$ of $\sim$0.25. 
Using the peak surface brightness and the length of the longest line of sight,
$2 \sqrt{2R\Delta R - (\Delta R)^2}$, we then derived a density in the 
shell of 40 cm$^{-3}$. On average, the diffuse gas of N44\,F probably 
possesses a density of a few tens cm$^{-3}$.

Using the Curtis Schmidt \ha\ image  and the flux calibration from 
\citet{ken}, we measured over the entire N44\,F nebula an \ha\ flux
of 3.6$\times$10$^{-11}$ erg~cm$^{-2}$~s$^{-1}$. For a distance of 
50 kpc, the \ha\ luminosity of N44\,F is then 1.4$\times$10$^{37}$ erg~s$^{-1}$,
after correcting for an extinction of E(B$-$V)=0.11~mag. This \ha\ 
luminosity requires an ionizing flux of $Q(H^0)$=1.0$\times$10$^{49}$ 
photons~s$^{-1}$. A second evaluation of the extinction correction 
has been made by comparing radio data and the Curtis Schmidt \ha\ image: 
a mean value E(B$-$V) of 0.28~mag for N44\,F, N44\,J, and N44\,M is found 
(J.\ R.\ Dickel, private communication), and this results in a $Q(H^0)$ 
of 1.4$\times$10$^{49}$ photons~s$^{-1}$ for N44\,F. This value suggests
that the total ionizing power of the star(s) within N44\,F is equivalent to
a single O6.5 V--O7.5 V star \citep{pan, vac, sdk}. 
This is compatible with the rather high \oiii/\ha\ ratio: the 
photoionization models of \citet{sta} for CoStar stellar atmosphere 
models suggest that the observed integrated \oiii/\ha\ ratio of N44\,F 
could be produced by the equivalent of an O8 V--O9 V star\footnote{We have
compared N44\,F, N44\,J, \& N44\,M to the closest model available in 
\citet{sta}, i.e.
a spherical nebula of density 10 cm$^{-3}$ and metallicity 0.25 Z$_{\odot}$, 
ionized by one single star.}.  The comparison to the photoionization models 
should however be interpreted with caution because the models do not assume 
a shell structure, which affects the excitation and ionization structure. 
The dominant spectral type could therefore be somewhat earlier than inferred 
from the modeled excitation.

\subsection{Pillar-like Dust Columns} 

The dark clouds of N44\,F exhibit new examples of the intriguing finger-
or pillar-like structures, similar to those reported in the Eagle Nebula 
\citep{hes96}. The most prominent two are embedded in the bright 
northwest region (see Figs.~\ref{n44f} and~\ref{pillars}). Several 
additional features of this type are present in the south and 
possibly in the southwest, but they are fainter. In Fig.~\ref{pillars}, 
we present the most striking of these structures in the \ha\ and 
\sii\ images as well as in the ratio map. The dust columns in N44\,F 
are 5\arcsec\ to 10\arcsec\ long and 1\arcsec\ to 2 \arcsec\ wide, 
corresponding to 1 to 2 pc long and 0.25 to 0.5 pc wide at the LMC's 
distance. These sizes
and morphologies are similar to those seen in 30 Doradus \citep{sco98} 
and M16 \citep{hes96}. Two of these columns harbor stars 
at their tips, as indicated by arrows in Fig.~\ref{pillars}. \citet{hes96} 
suggested that stars at the tips of pillars could be young stellar 
objects which were formed in the molecular cloud and finally emerge 
after the photoevaporation of most of their parent cloud.  However,
the \citet{hes96} model of pre-existing condensations in the molecular
cloud does not provide the unique conditions for producing 
photoionized pillars.  Hydrodynamical simulations by \citet{wil}
illustrate that these structures can be formed by a variety of initial
conditions in the structure with inhomogeneities in density and/or 
radiation field, but current observations do not yet permit 
discrimination among these different models.  It is therefore of great 
interest to identify additional examples of such objects.

Other important characteristics of these dust columns are their ionization 
structure and their density.  All of the dust columns point towards
the central ionizing source, i.e., a bright star or a small cluster 
between the center and the northwest edge of N44\,F. 
As can be seen in Fig.~\ref{ionstruc}, the 
\sii\ peak is separated from the \ha\ peak by approximately 0\farcs1, 
i.e., 0.024 pc, with \ha\ peaking slightly closer to the source of UV 
radiation which is photoevaporating the cloud. The shape of these 
profiles is very similar to those observed in the pillars of 30 Doradus,
or in ground-based images of M16 \citep{sco98}. Unfortunately, no WFPC2 
\oiii\ images of N44\,F are available, thus we can not undertake a complete 
analysis using photoionization models or constrain further the properties 
of the ionizing source. Following the method used in section 3.1, we estimate 
an rms density larger than 300 cm$^{-3}$ at the tip of one of the pillars. 
This density is compatible with that determined by \citet{sco98} in 30
Doradus, 700~cm$^{-3}$.  These pillars are thus photoevaporating
clouds, whose evaporated matter enhances the density and
brightness of the diffuse material seen in the northwest part of
the nebula.

\subsection{Kinematic Properties} 
The kinematic properties of N44\,F are derived from the high-dispersion
echelle observations of the \ha\ and \nii\ $\lambda$6583 lines.
The \nii\ line has a smaller thermal width than the \ha\ line,
thus it shows the line splitting better although it is noisier.
We have extracted \ha\ and \nii\ line profiles and fitted them with 
Gaussian components.  The resultant velocity-position plot (Fig. 
\ref{specn44}) shows that N44\,F is an expanding shell with a radius 
of 22\arcsec, coincident with the outer ring described in \S 3.1. Both 
lines indicate a shell expansion velocity of $\sim$12 \kms. 
An example of the \nii\ and \ha\ profiles are presented in Fig.~\ref{prof}. 
The approaching (blueshifted) component is fainter, suggesting a 
lower density. 
Further east, outside N44\,F, an additionnal blue component is
present, but it is probably associated with the complex filamentary
structure at the outskirts of N44 and is not directly related 
to N44\,F.

\subsection{Test of the Standard Bubble Model}
N44\,F has a classical ring-nebula morphology, and its kinematics confirms 
an expanding shell, most likely created by the wind of the early-type 
massive star(s) within this bubble. We can now test whether the 
observed expansion dynamics of this bubble are consistent with those 
expected from the standard, adiabatic bubble model \citep[e.g.,][]{Wetal77}.
With the observed radius $R$ in units of pc and expansion 
velocity $V$ in units of \kms, we can compute the dynamical timescale
$t_6$ in units of 10$^6$ yr using $t_6 = 0.6 R / V$, and further 
determine the ratio of stellar wind luminosity $L_{36}$ in units
of 10$^{36}$ ergs~s$^{-1}$ to ambient density $n_0$ in units
of H-atom~cm$^{-3}$, using $R=27\,(\frac{L_{36}}{n_0})^{1/5}\,t_6^{3/5}$.

For a radius of 5.3 pc and an expansion velocity of $\sim$12 \kms, 
N44\,F has a dynamical timescale of 2.7$\times$10$^5$ yr and a $L_{36}/n_0$ 
ratio of 0.016 ergs\,\,s$^{-1}$\,\,cm$^{3}$. If we assume that the shell 
consists of mass swept up from the volume, the ambient density, $n_0$, can 
be estimated by $n_0 = 3 n \frac{\Delta R}{R}$, where $n$ is the rms density 
of the shell, $\Delta R$ the shell thickness, and $R$ the shell radius. For
$n\sim$40 cm$^{-3}$ and $\frac{\Delta R}{R}=0.25$ (see Section 3.1), we derive 
an ambient density of $\sim$30 cm$^{-3}$ and a wind luminosity of 
$\sim$5$\times$10$^{35}$ ergs~s$^{-1}$. This wind luminosity is compatible 
with that expected from the spectral type determined in Section 3.1, i.e., 
an O7V star \citep{deJager88, Petal90}. One uncertainty in these estimates
is the shell thickness; however, even if the shell thickness is off by a 
factor of two, the predicted wind luminosity will still be compatible,
within the errors, with an O7V star.

Thus, the dynamics of N44\,F appear to be consistent with the standard
adiabatic model for bubble evolution, within uncertainties generated by
the complex interstellar environment.  This contrasts with the results
from superbubbles, which show evidence of retarded growth \citep[e.g.,][]{Oey96}.  To investigate further this discrepancy, it is 
worth considering an alternative bubble model.

The presence of dust pillars around N44\,F clearly indicates a highly 
inhomogeneous ISM. The photoevaporation of the dense
clumps may serve as an energy sink and govern the dynamic evolution 
of this \hii\ region.  Since photoevaporation processes take place in 
the early evolutionary stages of a bubble/superbubble, the energy 
loss by photoevaporation may explain the smaller-than-expected sizes 
of previously studied superbubbles and Wolf-Rayet ring nebulae (see \S 1).
Therefore, we will now consider models of bubbles formed in a cloudy 
medium by \citet{mckee}.
In their models, only bubbles blown by stars with weak winds will 
follow the Weaver et al.\ model, while stars with stronger winds
will produce bubbles whose expansion is governed by the dynamics 
of photoevaporating clouds.  They define a dimensionless wind luminosity 
$L_W^{\ast}\equiv {L_W}/({1.26\times10^{36} (S^2_{49}/n_m)^{1/3}})$ 
and a homogenization radius $R_h({\rm pc})\equiv 0.009\,\,t^{4/7}\, 
\left(\frac{S_{49}}{n^2_m}\right)^{1/7}$, where $L_W$ is the wind 
luminosity in erg~s$^{-1}$, $S_{49}$ the ionizing flux in
 photons~s$^{-1}$, $n_m$ the mean cloud density in cm$^{-3}$ and $t$ 
the dynamical timescale in years. Strong winds possess 
$L_W^{\ast} \gtrsim 1$ and blow bubbles of radius $\sim R_h$. 

Using the observed parameters of N44\,F (with $n_m$=$n_0$)
and assuming a stellar wind luminosity for an O7V star, we 
estimate $L_W^{\ast}\sim$1 and $R_h\sim$4.5~pc, close to the observed 
bubble's radius. This seems to suggest that the dynamics of 
photoevaporation actually governs the bubble's expansion. 
The agreement of the data with both models could be
a simple coincidence, or maybe the bubble has just encountered the cloudy
medium and begun the photoevaporation processes, in which case there may 
not be a large deviation from the Weaver et al. model.
Note, however, that all these conclusions hinge upon the validity
of the stellar wind luminosity, which is based on an assumed spectral
type of the central star.  Future study of the actual stellar content
of N44\,F is needed to place stronger constraints on the applicable
bubble models.

\section{Physical Structure of N44\,J and N44\,M}

\subsection{N44\,J} 
Situated in the northernmost part of the N44 complex, N44\,J is a
slightly elongated, ring-like nebula with a size of 
37\arcsec$\times$32\arcsec (or 9.0 pc$\times$7.8 pc). The 0.9~m \ha\ 
image shows a bright rim around dusty features at the northwest boundary. 
A bright central star is visible in the \ha\ and \sii\ images (see 
Fig.~\ref{n44jm}). In \ha\ and \sii, the nebula exhibits modest 
limb-brightening, with the edge-to-center brightness ratio $<$2, 
while its central part 
displays a rather uniform surface brightness.  The \sii/\ha\ ratio is 
very similar to that of N44\,F, ranging from 0.25 to 0.4. In contrast, 
the \oiii\ image of N44\,J shows emission only near the central star. The 
\oiii/\ha\ ratio is 0.4 near the star, and only 0.05--0.08 elsewhere, 
significantly lower than the corresponding value in N44\,F. The integrated 
\oiii/\ha\ ratio of N44\,J is about 0.12.

The observed \ha\ flux of N44\,J is 5.7$\times$10$^{-12}$ 
erg~cm$^{-2}$~s$^{-1}$. This requires an ionizing flux $Q(H^0)$ 
of 1.6$\times$10$^{48}$ photons~s$^{-1}$ when adopting E(B$-$V)=0.11~mag,
or 2.3$\times$10$^{48}$ photons~s$^{-1}$ for E(B$-$V)=0.28~mag.
These values are expected for an O9V--B0V type ionizing star 
\citep{pan, vac, sdk}, which is fully consistent with that suggested by 
the \oiii/\ha\ ratio of the nebula \citep{sta}. 
Assuming a spherical geometry, we derive an rms electron density 
of $\sim$25 cm$^{-3}$, slightly less dense than N44\,F. The ionized gas 
mass of N44\,J is then $\sim$200 M$_{\odot}$.

The echelle spectrum of N44\,J shows a kinematically quiescent \hii\ 
region. The centroids of the \ha\ and \nii\ lines are centered at 
292 \kms, with an rms variation of 3 \kms\ (see Figs.~\ref{specn44} and 
\ref{prof}). 
No line splitting is seen. The observed FWHMs are 27$\pm$1 \kms\ for 
\ha\ and 16$\pm$2 \kms\ for \nii. These observed line widths imply a 
turbulent FWHM of about 10 \kms\ for \ha\ and 8 \kms\ for \nii\ if we 
take into account the instrumental profile and the thermal broadening 
at 10$^4$\,K. The turbulence can thus explain the small variations of 
the centroid velocity. The difference between the turbulence widths is 
probably caused by a difference in the emission regions of the two lines. 
If N44\,J is actually expanding, the expansion velocity should be very small,
$<$8 \kms. For such small expansion velocity, no strong compression is 
expected, hence no sharp filaments should be seen.

Finally, we may compare the observations with the standard, adiabatic
bubble model. As the density in the ring-like nebula is a few tens cm$^{-3}$, 
we will assume in our calculations a density of 10 cm$^{-3}$ in the ISM. 
For an expected wind luminosity of $\sim$3$\times$10$^{34}$ erg\,s$^{-1}$ 
for an O9.5V star \citep{deJager88,Petal90} and with the observed 17\farcs5 
radius of N44\,J, the \citet{Wetal77} model predicts an expansion velocity
of about 8 \kms\ for N44\,J. This expected expansion velocity is just at the limit implied by the velocity widths of the \ha\ and \nii\ lines.  
Raising the ambient density by a factor of 2 can lower the 
expected expansion velocity to 6.5 \kms, which is more consistent with 
the observed limit.  We conclude that the observed bubble dynamics of 
N44\,J does not greatly contradict the model predictions. 

\subsection{N44\,M} 
On the southeast part of N44, another small ring-like nebula is
present: N44\,M. It is the faintest nebula of our sample. It shows a
ring with a radius of $\sim$22\arcsec\ (or 5.3 pc), but its north
boundary is not well defined (see Fig.~\ref{n44jm}). The sharp southwest 
boundary suggests the presence of dust clouds. Several bright stars appear 
in its interior. N44\,M possesses an excitation structure very similar to 
N44\,J. The \sii/\ha\ ratio of N44\,M is comparable to that in N44\,F, but the 
\oiii/\ha\ ratio of N44\,M is much lower, ranging from 0.07 to 0.16 in the 
nebula and rising to 0.4 near the stars. The integrated \oiii/\ha\ ratio 
of N44\,M is about 0.18.

With an observed \ha\ flux of 5.3$\times$10$^{-12}$ erg\,cm$^{-2}$\,s$^{-1}$
and the two evaluations of the extinction, the required ionizing flux 
$Q(H^0)$ is 1.5--2.1$\times$10$^{48}$ photons\,\,s$^{-1}$. This 
suggests an O9-B0V type ionizing star \citep{pan, vac, sdk}, which is again 
compatible with the \oiii/\ha\ ratio \citep{sta}. If we assume 
a spherical geometry, we find that the rms density is about $\sim$17 
cm$^{-3}$ and the estimated mass of the nebula $\sim$275 M$_{\odot}$. 

The echelle spectrum of N44\,M shows only one component, centered at 298 
\kms\ with an rms variation of 1 \kms, and a FWHM of 28$\pm$1 \kms\ in the 
\ha\ line and 20$\pm$4 \kms\ in the \nii\ line (see Figs.~\ref{specn44} and 
\ref{prof}).  The turbulent FWHM are 
then 12 \kms\ for \ha\ and 14 \kms\ for \nii.  With an observed 
radius of $\sim$22\arcsec, a density of 10 cm$^{-3}$ for the ISM and a 
O9.5V ionizing star (see Section 4.1.), the standard bubble model predicts 
an expansion velocity of $\sim$7 \kms\ for N44\,M, which appears consistent 
with the upper limit of the echelle observations.

The kinematic characteristics of N44\,J and N44\,M are very similar, despite 
the somewhat different morphology. Neither \hii\ region contains
shells expanding at velocities greater than $\sim$8 \kms. In the future, 
additional observations may enable a better comparison of these nebulae with 
the standard model. Precise photometry and/or stellar spectroscopy would
give accurate informations about the stellar content of these nebulae,
while nebular spectra with higher resolution should more strongly constrain 
the kinematics of N44\,J and N44\,M. Thus, improved data is needed to
more stringently test the standard model for N44\,J and N44\,M.

\section{Conclusion}

Massive stars inject a large amount of energy into the ISM
through their ionizing radiation, fast stellar winds, and SNe.  The 
stellar winds, which dominate the early mechanical feedback, are able
to sweep the ambient medium into an expanding shell and create a
cavity around the parent stars.  \hii\ regions ionized by massive
stars are therefore thought to evolve from compact \hii\ regions to
ring-like nebulae.

We have studied three candidate wind-blown bubbles in the N44
complex: N44\,F, N44\,J, and N44\,M. They all have diameters of $\sim$10~pc. 
Their ionizing fluxes and \oiii/\ha\ ratios suggest
that their ionizing stars have spectral types of roughly O7V, O9.5V and
O9.5V, respectively. N44\,F has an expansion velocity of 12 \kms,
while the other two have expansion velocities $<$ 8\kms.
The stellar wind luminosity implied by the ionizing stars
and the observed bubble dynamics appear consistent with that
expected in the standard, adiabatic bubble model within the
limits of observational uncertainties.  Similar results were
found by \citet{OM94} for two bubbles around single O star
in M33.  This is in sharp contrast to the anomalous kinematics
seen in superbubbles and Wolf-Rayet bubbles.
We caution, however, that our observed expansion velocities may
not provide the most stringent constraints when comparing with
models, because the expected expansion velocities of N44\,J
and N44\,M are comparable to or smaller than the convolution of the
isothermal sound velocity ($\sim$10 \kms) and the instrumental
FWHM (13 \kms).  Deeper, higher-resolution spectroscopic
observations of the \nii\ or \oiii\ lines are needed to pin down
the expansion velocity for a more definitive test of models.

We also report the discovery of several dust pillars in N44\,F, which
are similar to those found in the Eagle Nebula \citep{hes96} and 30 
Doradus \citep{sco98}.  These structures are indicative of the
evaporation of molecular clouds by the ionizing source.  The examples 
in N44\,F are especially well-suited to investigating the process and 
role of photoevaporation in star formation and shell evolution, because 
of their location within such a well-defined, wind-blown bubble.  We 
found that photoevaporation, as evidenced by the pillar structure, may 
be dynamically important for the shell evolution in N44\,F, slowing down 
the expanding shell and inhibiting its growth. These photoevaporation 
processes may thus undermine the seeming agreement between the observed 
bubble dynamics and the simple adiabatic bubble model.

\acknowledgments 

We thank Charles Danforth for assisting in the echelle observations and 
Rosie Chen for help in the {\it HST} WFPC2 image reduction. This work has 
been supported by the NASA grant STScIGO-6698.01-95A. Y.N. acknowledges 
support from the F.N.R.S., contracts P4/05 and P5/36 `P\^ole d'attraction 
Interuniversitaire' (SSTC-Belgium) and the PRODEX XMM-OM and Integral Projects.
R.C.S. acknowledges the support of the Dean B. McLaughlin fellowship at 
the University of Michigan and NSF grant AST 95-30747.

\clearpage

\begin{figure} 
\caption{ CTIO 0.9~m \ha\ (+\nii) CCD image of N44. The nebular components
defined by Henize (1956) and the OB associations cataloged by Lucke 
\& Hodge (1970) are marked. \label{n44tot}} 
\end{figure} 

\begin{figure}
\caption{ Ground-based images of N44\,F (left), N44\,J (middle), and N44\,M 
(right). The first row contains \ha\ (+\nii) images taken with the CTIO 0.9~m 
telescope. The second, third, and fourth rows are Curtis 
Schmidt images taken in \ha\ (+\nii), \oiii, and \sii\ filters, 
respectively. The field-of-view (FOV) of each image is 92\arcsec$\times$92\arcsec.  North is up and east to the left.
\label{n44jm}}
\end{figure} 

\begin{figure} 
\caption{{\it HST} WFPC2 \ha\ (+\nii) image (left), \sii\ image (middle) and
\sii/\ha\ ratio map (right) of N44\,F. The rectangular boxes show
the position of the close-up images presented in Fig. 4.
The bars below the images show the greyscales for surface brightness 
(in units of erg cm$^{-2}$ s$^{-1}$ arcsec$^{-2}$) or line ratio. 
North is up and east to the left.
\label{n44f} }
\end{figure}

\begin{figure} 
\caption{ Close-up images of dust columns seen in N44\,F through \ha,
\sii, and \sii/\ha\ (for exact position, see Fig. \ref{n44f}). Left: 
dust columns situated at the west rim of N44\,F 
(FOV=18\arcsec$\times$18\arcsec). Right: dust columns situated at the 
south rim of N44\,F (FOV=14\arcsec$\times$14\arcsec). The arrows point 
to the stars at the tips of the pillar-like dust columns.
\label{pillars}} 
\end{figure} 

\begin{figure} 
\caption{ Left: \ha\ image of the ``pillars'' situated at the west part of N44\,F.
The rectangular box and the arrow show the position and direction of the cut 
presented at the right. Right: Ionization structure along the rectangular 
aperture. \sii\ and \ha\ intensities are normalized to the peak, while 
the \sii/\ha\ ratio is the actual value. Note that the \ha\ and \sii\ peaks 
are separated by one pixel, i.e. 0\farcs1.
\label{ionstruc}} 
\end{figure} 

\begin{figure}
\caption{ Top panel: Echellogram of the \nii\ $\lambda$6583 line for N44\,F. 
Bottom panels: 
Radial velocity-position plots of the echelle slit positions in N44\,F, 
N44\,J, and N44\,M for both \ha\ and \nii\ lines.  The radial velocities 
are heliocentric.  The \ha\ velocity components are plotted in filled 
symbols, and the \nii\ in open ones. The position axis is defined with 
the center of the nebula at the origin (0\arcsec) and increasing towards 
west. The dot-dashed line in each plot indicates the range covered by
the nebula. 
Arrows show the positions of the profiles presented in Fig. \ref{prof}.
\label{specn44}}
\end{figure} 

\begin{figure}
\caption{ \ha\ and \nii\ profiles at the positions indicated in Fig. 
\ref{specn44}.
\label{prof}}
\end{figure} 

\end{document}